\documentclass[aps,prl,superscriptaddress,twocolumn]{revtex4}
\usepackage{graphicx}
\usepackage{times}
\usepackage{amsmath}
\usepackage{amsthm}
\usepackage{amssymb}
\usepackage{amsbsy}
\newcommand{\abs}[1]{\vert #1 \vert}
\newcommand{\bra}[1]{\langle #1 \vert}
\newcommand{\ket}[1]{\vert #1 \rangle}

\newcommand{\C}{\hat{c}}
\newcommand{\D}{\hat{d}}

\begin{document}
\title{An experimentally accessible quality factor for Majorana wires}
\author{David J. Clarke}
\affiliation{Condensed Matter Theory Center, Department of Physics, University of Maryland, College Park}
\affiliation{Joint Quantum Institute, University of Maryland, College Park}
\affiliation{Station Q Maryland}
\begin{abstract}
    Spin-orbit coupled semiconducting nanowires with proximity-induced superconductivity are expected to host Majorana zero modes at their endpoints when a sufficiently strong magnetic field is applied. The resulting phase would be a one-dimensional topological superconductor. However, while a variety of experiments have been performed observing a zero bias conductance peak (suggestive of Majorana zero modes), the topological nature of these physical systems is still a subject of debate. Here we suggest a quantitative test of the degree to which a system displaying a zero bias peak may be considered topological. The experiment is similar to previous measurements of conductance, but is performed with the aid of a quantum dot at the wire's end.  We arrive at the surprising result that the non-local nature of the topological system may be identified through a local measurement.
\end{abstract}
\maketitle
    Non-Abelian anyons, quasiparticle excitations in two-dimensional systems that enact non-commuting unitary transformations on the ground state of a system when exchanged, are a fascinating consequence of low-dimensional physics, and are expected to provide a significant advantage in the field of quantum computation\cite{Nayak08, Alicea12a, Leijnse12, Beenakker2013a, Stanescu13, DasSarma15, Elliott15}. Non-Abelian statistics of this sort has yet to be demonstrated definitively, but recent years have shown a flurry of experimental \cite{Mourik12,Das12,Deng12,Rokhinson12,Finck12,Churchill13,Chang15,Zhang16,LeeEJH14,Albrecht16,Deng16} and theoretical \cite{Kitaev01,Sau10a,Lutchyn10,Oreg10,Fu10, Fidkowski12, adagideli2014effects, DasSarma16, liu2012zero, LinC12, Moore16,Alicea11,Sau11b,Halperin11,Clarke11b,Clarke16a,Aasen16,Hyart13,Clarke16,Plugge16,Karzig16,Vijay16} progress in the development of the simplest type of non-Abelian excitation, defects binding real fermionic zero modes called Majorana zero modes or MZMs. Some of the most exciting of these results have been the observation of zero bias conductance peaks in spin-orbit coupled semiconducting nanowires with proximity-induced superconductivity. Such systems are expected to host Majorana zero modes at the ends of a region of an effective spinless p-wave `topological' superconducting region \cite{Sau10a,Lutchyn10,Oreg10,Kitaev01}.  While experiments thus far are extremely suggestive of Majorana physics and tend to defy explanations of a non-topological nature (\emph{e.g.} Kondo physics) \cite{Fidkowski12, adagideli2014effects, DasSarma16, liu2012zero, LinC12, Moore16, LeeEJH14}, the observation of a zero-bias conductance peak is itself only a necessary and not a sufficient condition for determining that a system is behaving as a topological superconductor. For this reason, a variety of experiments have been proposed for probing the non-Abelian nature of the Majorana defects as an indication of topological physics \cite{Alicea11,Sau11b,Halperin11,Clarke11b,Clarke16a} and as a stepping stone toward quantum computation \cite{Hyart13,Clarke16,Aasen16,Plugge16,Karzig16,Vijay16}. These proposals have in common that they involve non-local probes of the system (\emph{e.g.} measurement of the fermion parity of an entire wire segment). It is a unique feature of the measurement that we propose here that it is both local and sufficient to determine whether a system is behaving topologically.

    \begin{figure}
        \includegraphics[trim=0cm 9cm 0cm 1cm,clip, width=\columnwidth]{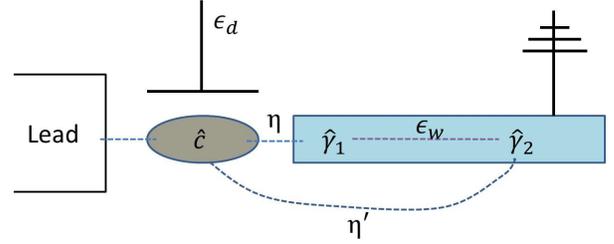}
        \caption{Schematic diagram of the system under consideration. Conductance spectroscopy is performed on a wire containing a single low-lying fermionic level $\D=\frac12(\gamma_1+i\gamma_2)$ by coupling to a lead at voltage $V_\mathrm{bias}$ relative to ground through a quantum dot with a single fermionic level with annihilation operator $\C$. The energy difference $\epsilon_d$ between the occupied and unoccupied states of that dot is set by a side-gate voltage and varied to perform the experiment suggested here. The dot-wire system is characterized by three additional parameters: the couplings $\eta$ and $\eta'$ from the dot to the Majorana modes $\gamma_1$ and $\gamma_2$ respectively, and the Majorana hybridization $\epsilon_w$ that sets the energy difference between occupied and unoccupied states of the fermionic mode in the wire. If the wire is in the topological superconducting phase, both $\eta'$ and $\epsilon_w$ should be $\sim 0$. A similar differential conductance experiment without the dot present can only access the value of $\epsilon_w$. With the dot, all three parameters may be measured independently. In particular, one can make measure of $q=1-\abs{\eta}/\abs{\eta'}$, a `topological quality factor' quantifying the extent to which one may expect non-Abelian behavior from the Majorana bound states in the wire.   }
        \label{Fig:Setup}
    \end{figure}

    The reason that observation of a zero-bias conductance peak is not sufficient to determine that a nanowire is in a    topological phase is somewhat subtle. When the lead probing the system is a normal metal, a zero-bias peak in conductance does indicate that there is at least one pair of energy levels in the system that differ in quantum numbers by an electron and that are degenerate to within the resolution of the experiment. A zero-bias peak in conductance is therefore sufficient to determine that there exists a fermionic mode $\D$ in the nanowire at zero energy (within some bound given by the peak width). Such a degeneracy is predicted in the topological system\cite{Nayak08, Alicea12a, Leijnse12, Beenakker2013a, Stanescu13, DasSarma15, Elliott15}, where the fermionic mode is shared among two Majorana zero modes separated by the length of the wire. (This separation is the source of the `topological protection' of the degeneracy and the utility for quantum information applications \cite{Nayak08, Alicea12a, Leijnse12, Beenakker2013a, Stanescu13, DasSarma15, Elliott15}.) However, it is important to realize that by taking real and imaginary parts of \emph{any} fermionic zero mode, that mode may be formally divided into two Majorana modes $\gamma_1$ and $\gamma_2$ at zero energy (as $2\D=i\gamma_1+\gamma_2$, where $\gamma_1^2=\gamma_2^2=1$). This alone does not suffice for the system to be topological. The hallmark of the topological phase is that these two Majorana modes are separated in space, which cannot be determined from an examination of the energy of the fermionic mode in and of itself (although it may be strongly suggested if the ground state energy splitting becomes exponentially small as the wire length is increased) \cite{Nayak08, Alicea12a, Leijnse12, Beenakker2013a, Stanescu13, DasSarma15, Elliott15,Albrecht16}. Rather, to determine that a system is topological, one must show that each of these Majorana modes is localized to a different position, and in particular that each mode may only couple to operators near that position.

    Here, we show that a modified transport experiment interposing a quantum dot between the lead and the nanowire can measure separately the coupling of the dot to each of the two Majorana bound states that make up the fermionic mode in the wire. If the wire is topological, the dot should only couple to the Majorana bound state nearest the wire end next to the dot. We therefore introduce a `topological quality factor' given by one minus the ratio of the two couplings. This quality factor is 1 when the system is purely topological and zero for an electronic (non-superconducting) bound state. A quality factor near 1 is essential for the observation of non-Abelian behavior from Majorana systems in future experiments \cite{Clarke16a}.

\begin{figure}
        \includegraphics[width=.49\columnwidth]{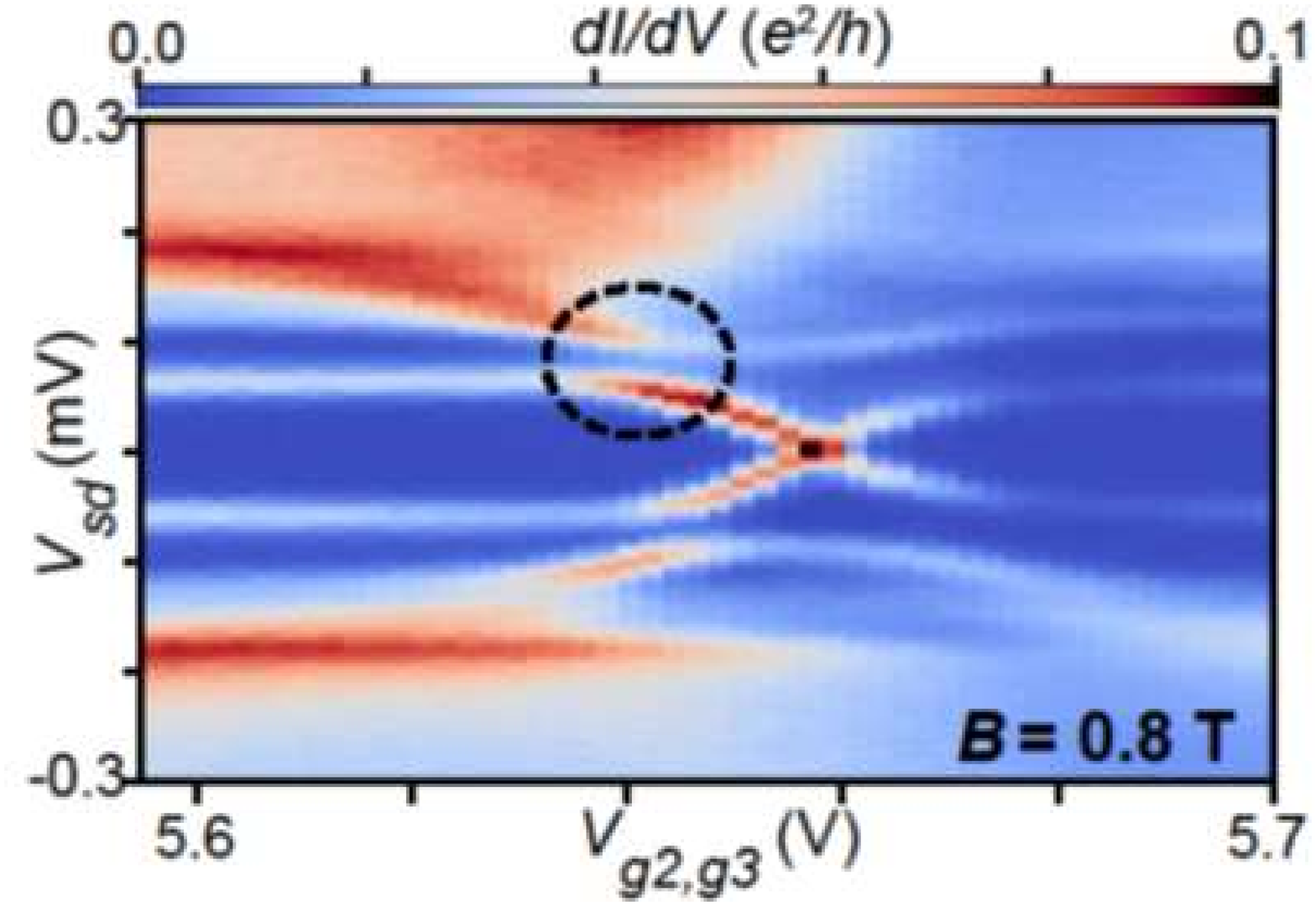}\includegraphics[width=.49\columnwidth]{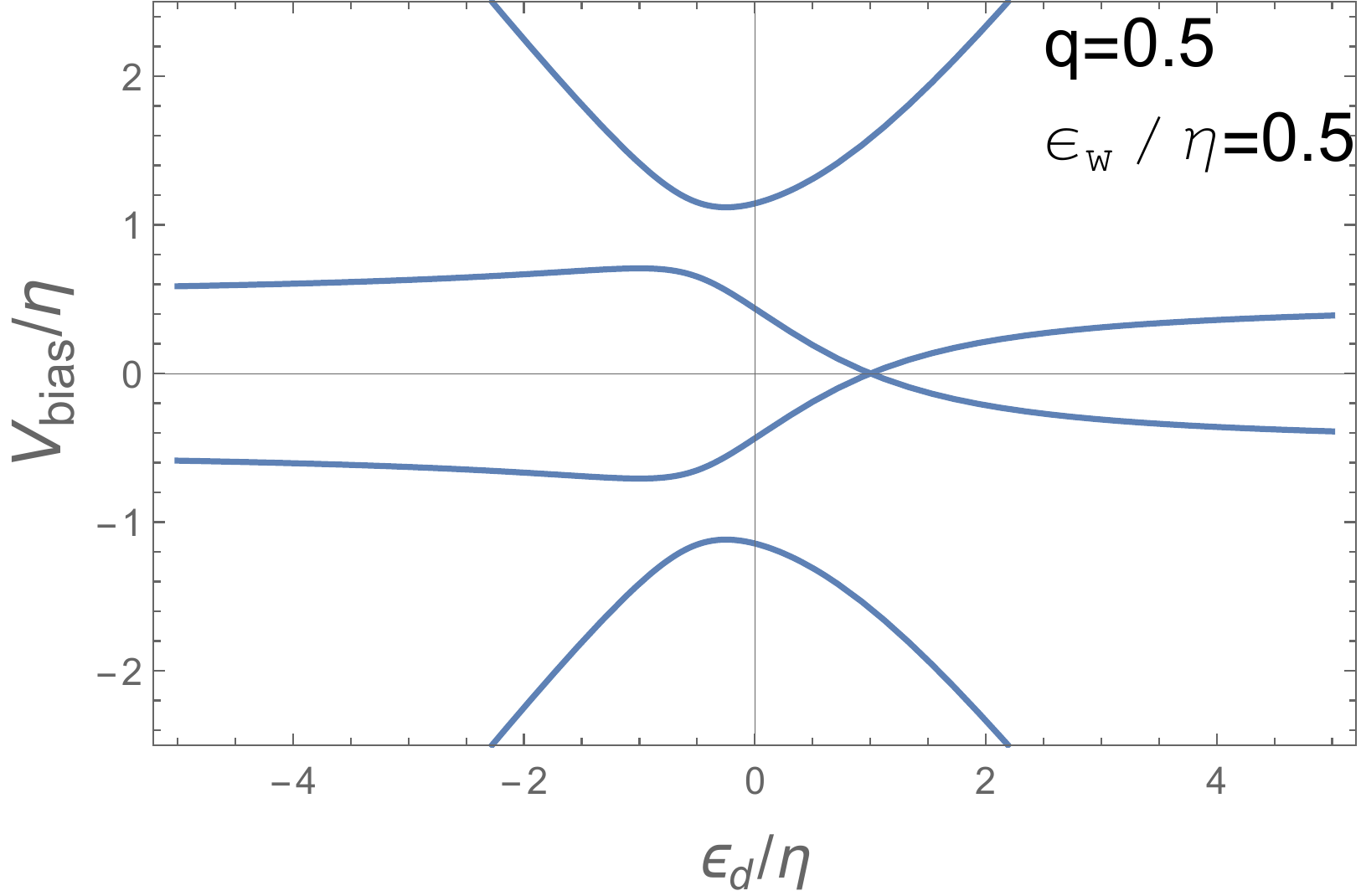}
        \includegraphics[width=.49\columnwidth]{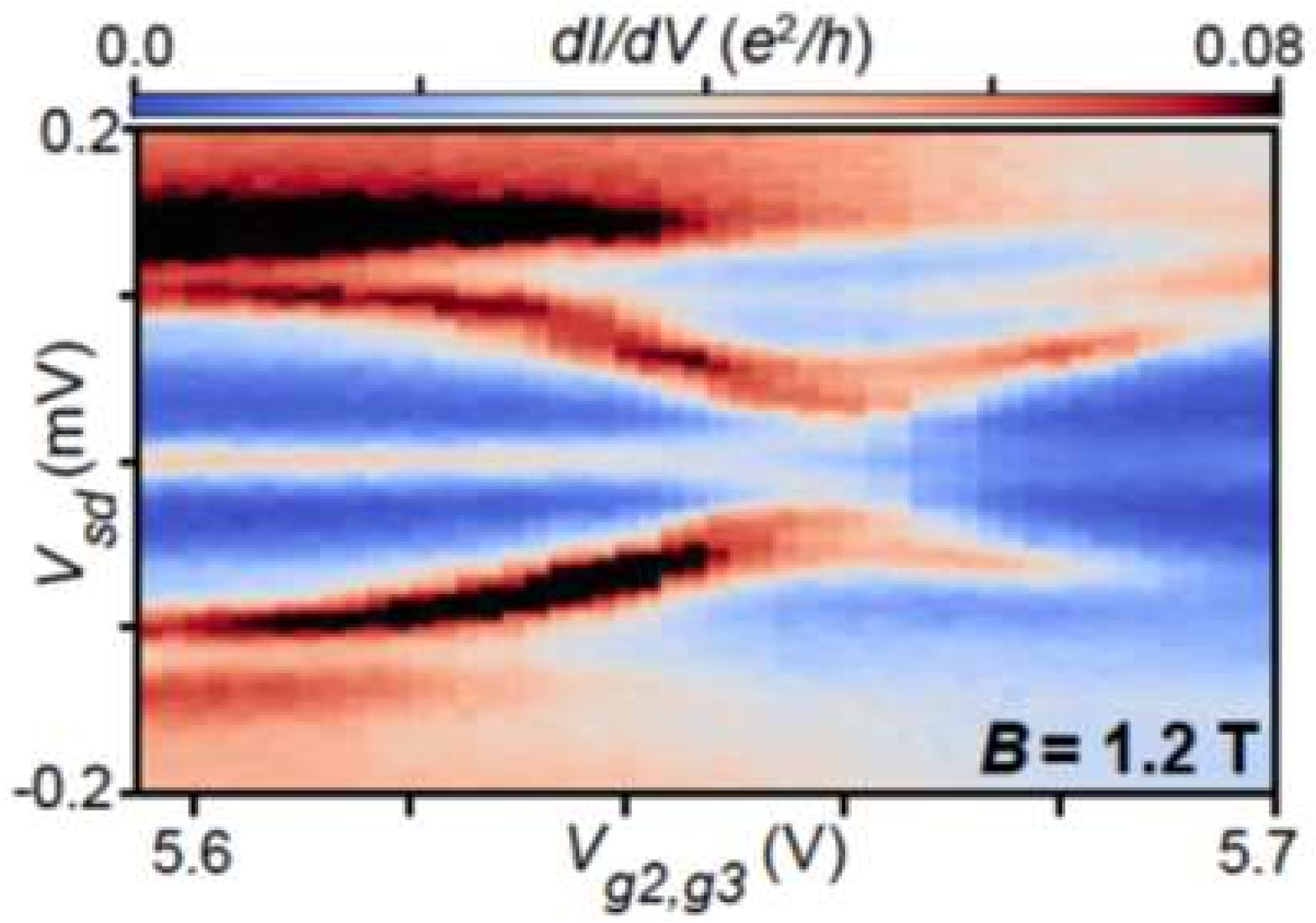}\includegraphics[width=.49\columnwidth]{Figq5}
        \caption{Comparison of differential conductance color plots from Deng \emph{et al.} \cite{Deng16} (left) with transition plots from our simple model (right). Here we plot the transitions for a quality factor of $q=0.5$ and two values of $\epsilon_w/\eta$ that lead to qualitatively similar traces. The circled point in the Deng \emph{et al.} data (top left) indicates an anticrossing between dot and wire states that is also present within our simple model (top right). In fact, the behavior is sufficiently alike that more exact values for the model parameters might be extracted from similar data in a systematic study.}
        \label{Fig:Deng}
    \end{figure}

    The quality factor may be obtained from differential conductance measurements by tuning the quantum dot through a resonance. One important property of the experiment we suggest is that it is eminently realizable in present day experimental systems. In fact, a version of this experiment has already been performed in Ref.~\onlinecite{Deng16}, though no systematic investigation of the quality factor was performed. We conclude our paper with a comparison of the conductance spectra in that paper with those predicted by the simple model presented here (See Fig.~\ref{Fig:Deng}). The qualitative similarly of the data and the results of our model should allow the effective extraction of the model parameters ($q$ in particular) from experiments on similar systems.

    \emph{The Model}: The bulk of this paper is devoted to the analysis of a simple model of the dot-wire system shown in Fig.~\ref{Fig:Setup}. We describe the quantum dot by a single fermionic mode $\C$ that may be occupied or unoccupied. The fermionic mode $\D$ in the wire is split into two Majorana modes $\gamma_1=-i(\D-\D^\dagger)$ and $\gamma_2=\D+\D^\dagger$. These modes are not assumed to be of topological origin. We note that charge is only conserved $\mod 2$ within the superconducting nanowire, so the most general Hamiltonian coupling the dot to the wire is given by
    \begin{equation}
      H=\epsilon_d\C^\dagger\C-\frac{i}{2}\epsilon_w\gamma_1\gamma_2 \C+\left(i\frac{\eta}{2}\C^\dagger\gamma_1 +\frac{\eta'}{2}\C^\dagger\gamma_2+\mathrm{h.c.}\right)
    \end{equation}
    Here the couplings $\eta$ and $\eta'$ may be taken to be real without loss of generality via gauge transformations on $\C$ and $\D$. We may express this Hamiltonian in the occupation basis for the fermionic modes on the dot and the wire. We have
    \begin{eqnarray}\label{eq:simple}
        H&=&\left(\begin{array}{c} \ket{01}\\\ket{10}\end{array}\right)^T
            \left(\begin{array}{cc}
                \epsilon_w & (\eta+\eta')/2 \\
                (\eta+\eta')/2  & \epsilon_d \\
            \end{array}\right)\left(\begin{array}{c} \bra{01}\\\bra{10}\end{array}\right)\nonumber\\
        &+&\left(\begin{array}{c} \ket{00}\\\ket{11}\end{array}\right)^T
            \left(\begin{array}{cc}
                0 &  (\eta-\eta')/2\\
                (\eta-\eta')/2 & \epsilon_w+\epsilon_d\\
            \end{array}\right)\left(\begin{array}{c} \bra{00}\\\bra{11}\end{array}\right)\nonumber\\
    \end{eqnarray}
    Note that we ignore any Coulomb interaction (such an interaction would add a term $E_C\ket{11}\bra{11}$ to the Hamiltonian).
    Again without loss of generality we take $\abs{\eta}>\abs{\eta'}>0$ and define the quality factor $q=1-\abs{\eta'}/\abs{\eta}$. 

    \begin{figure}
      \includegraphics[width=.49\columnwidth]{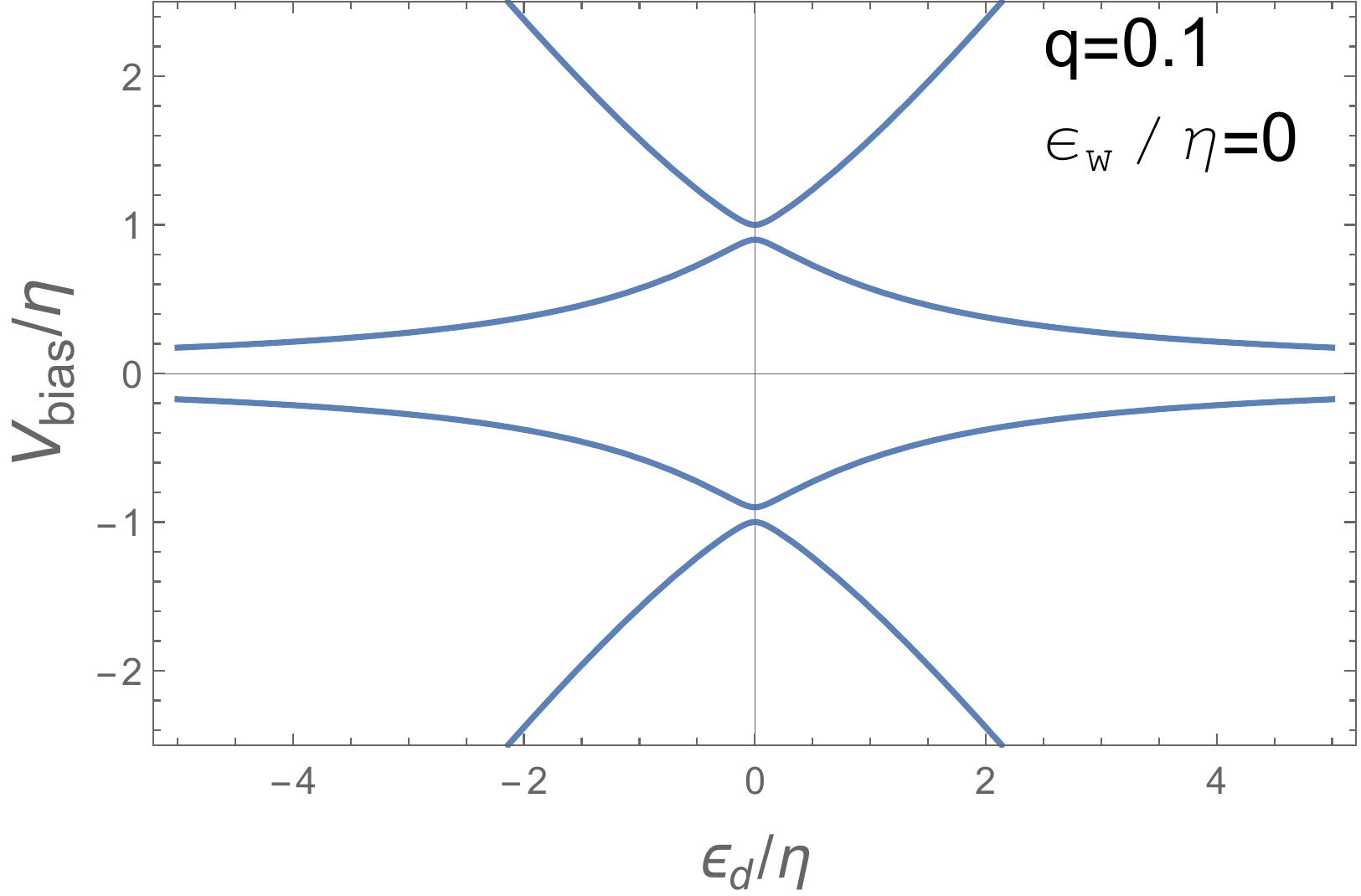}\includegraphics[width=.49\columnwidth]{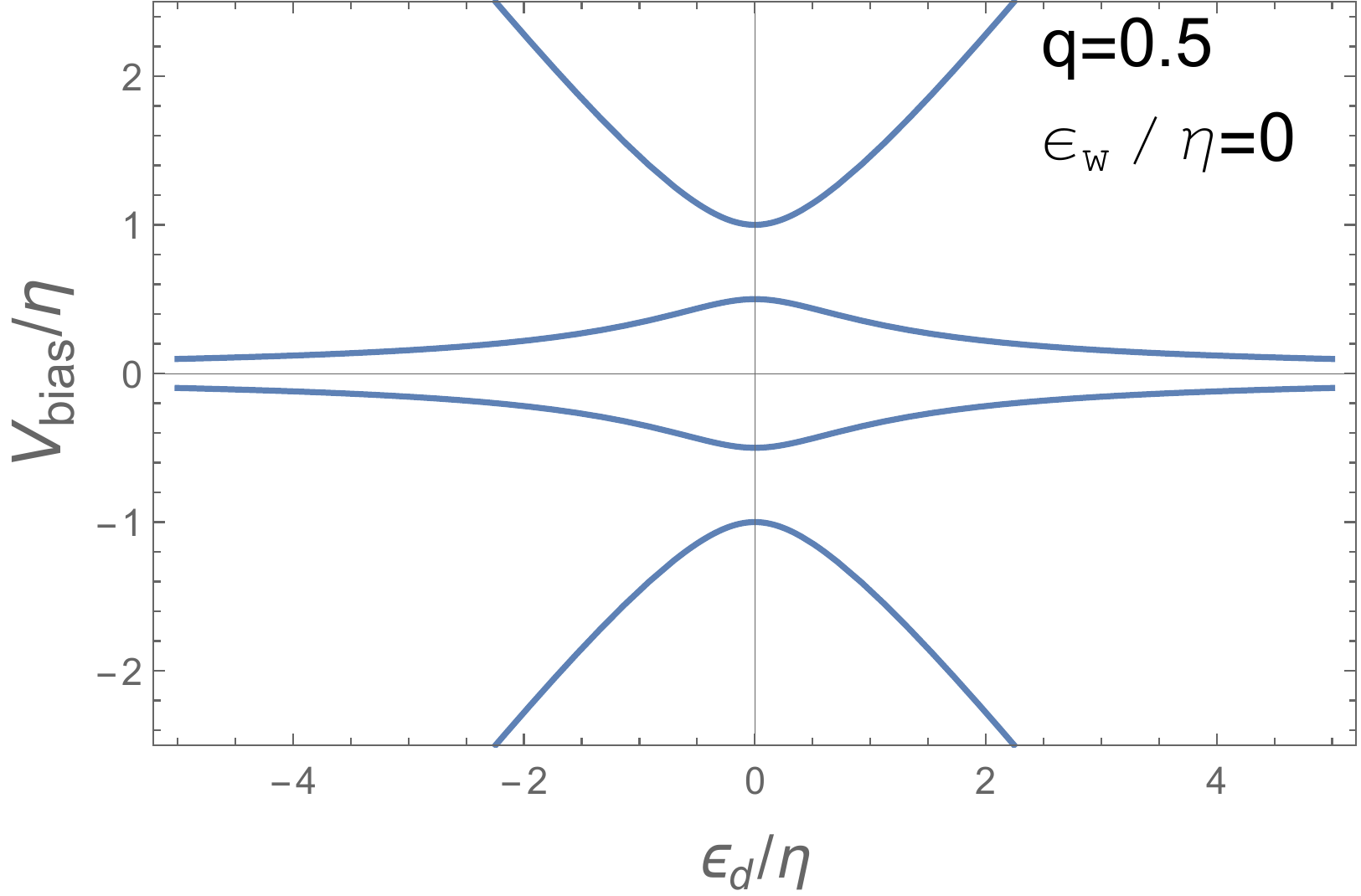}
      \includegraphics[width=.49\columnwidth]{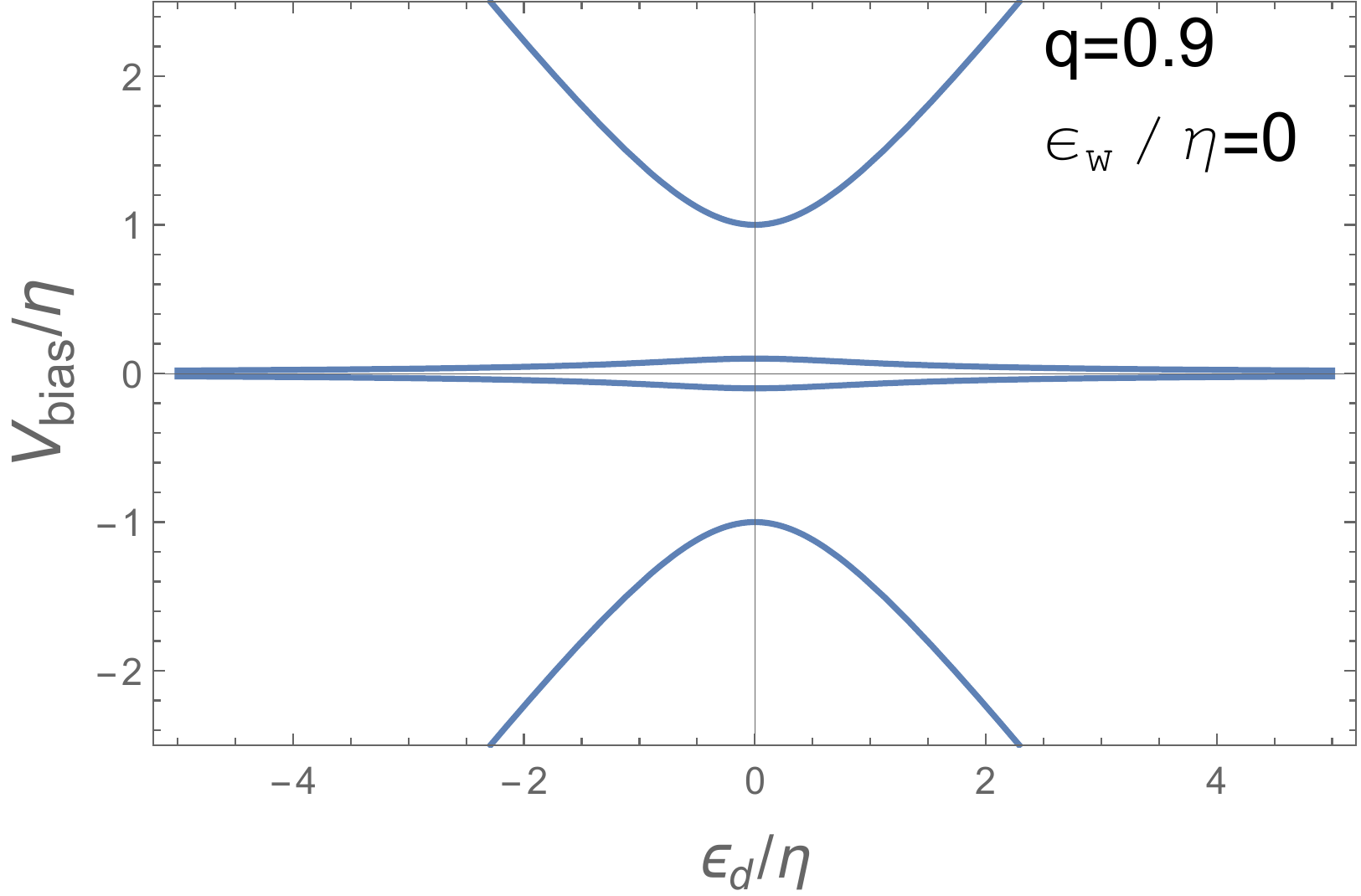}\includegraphics[width=.49\columnwidth]{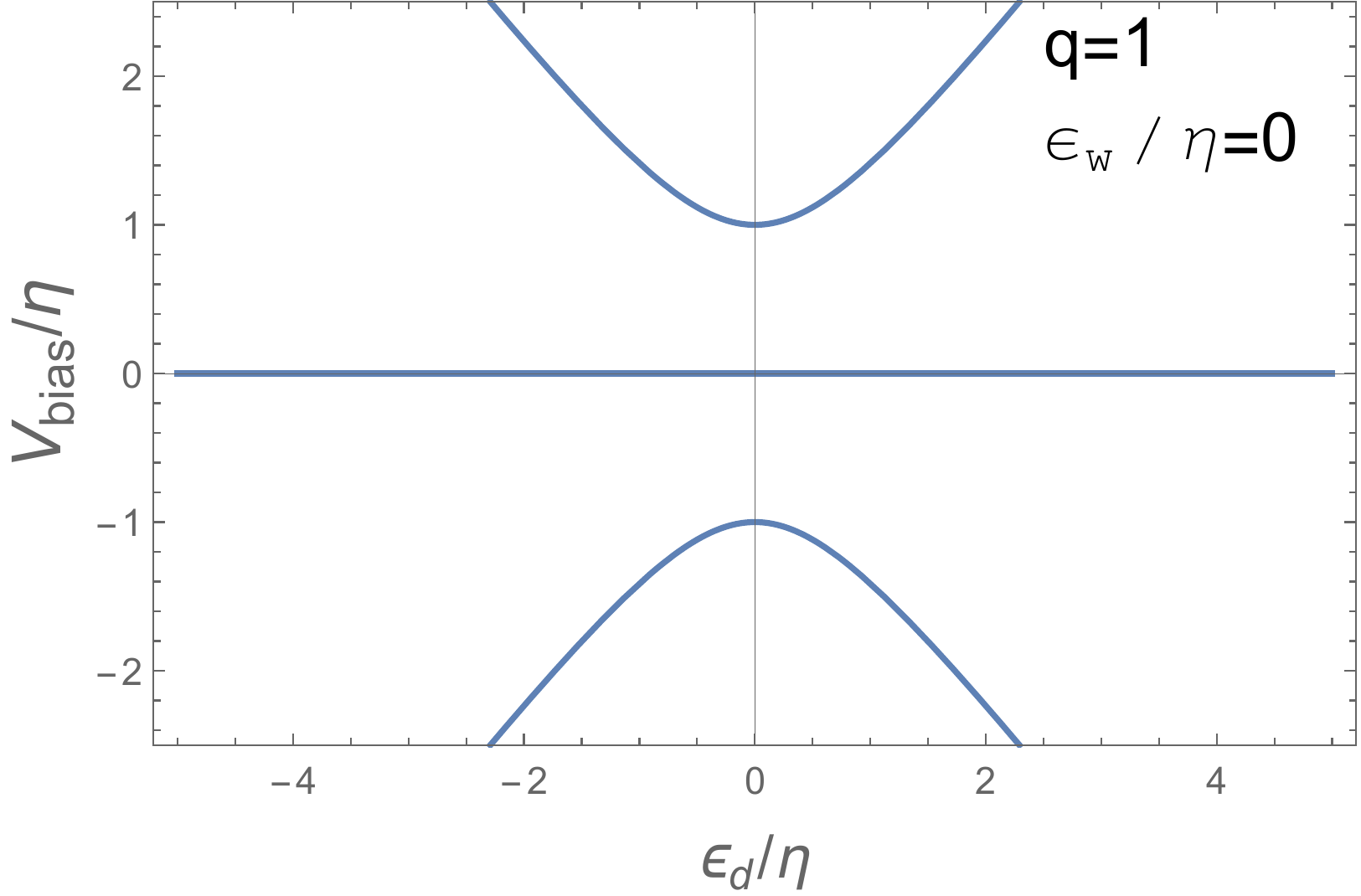}
      \caption{Expected low-energy spectrum near the dot resonance for various values of the topological quality factor $q$ and in the absence of hybridization between the two Majorana modes. The distance between the outer two mode peaks at resonance is twice the `topological' coupling $\eta$. The distance between the two inner mode peaks at resonance is twice the `non-topological' coupling $\eta'$. For a perfectly topological system ($q=1, \epsilon_w=0$), the zero-bias conductance peak does not split as it passes through the dot resonance.}
      \label{Fig:q}
    \end{figure}

    \emph{Measurement} We note that the Hamiltonian above is split into two subsectors with opposite fermion parity. A typical transport experiment measuring conductance through the dot into the superconducting wire has conductance peaks whenever the bias voltage is on resonance with transitions between the ground state of the dot-wire system and an excited state in the opposite parity sector. In the absence of a significant Coulomb interaction between the wires, the transition spacings available from each of the four possible ground states actually all coincide~\footnote{This degeneracy occurs because the trace of the Hamiltonian in each of the two parity sectors is equal.}. These conductance peaks will therefore occur at the values
    \begin{eqnarray}
     V_\mathrm{bias}&=&\pm\frac{1}{2}\sqrt{(\epsilon_w-\epsilon_d)^2+(\eta+\eta')^2}\nonumber\\
     &&\pm\frac{1}{2}\sqrt{(\epsilon_w+\epsilon_d)^2+(\eta-\eta')^2}.
    \end{eqnarray}

    If the dot is well off resonance ($\abs{\epsilon_d}\gg \eta,\eta'$), then the transitions between the lowest levels provides a direct measure of $\epsilon_w$, the `Majorana hybridization.' This measurement has been the focus of much of the experimental effort thus far, which finds $\epsilon_w\sim 0$ over a wide range of parameters \cite{Mourik12,Das12,Deng12,Rokhinson12,Finck12,Churchill13,Chang15,Zhang16,Albrecht16,Deng16}.

    However, by tuning the quantum dot to resonance \cite{Deng16}, we may achieve a direct measurement of the couplings between the dot and the wire. In the ideal (topological) case, only one of the Majorana modes in the wire is coupled to the dot ($\eta'=0$, or $q=1$). In Fig.~\ref{Fig:q} we plot the transition spectrum of the dot-wire system for $\epsilon_w=0$ and various values of $q$. Near resonance, two conductance peaks break off from the continuum and have an avoided crossing at $\epsilon_d=0$. Their point of closest approach gives a measurement of $\eta$, the larger of the two Majorana couplings (the distance between the peaks is $2\eta$). Meanwhile, the states that are degenerate at large $\epsilon_d$ split near the dot resonance by $2\eta'$.

    \emph{Comment on locality and Ising theory}:
    It is somewhat remarkable that such a simple experiment, measuring effectively local properties of the hybridization between the dot and wire states, can determine the non-local nature of the fermionic mode on the wire that is shared between the two Majoranas. It is somewhat revealing to approach this problem from the point of view of the Ising anyon theory that is (approximately) realized by Majorana zero modes \cite{Nayak08,Alicea12a,Leijnse12,Beenakker2013a,Stanescu13,DasSarma15,Elliott15}. This theory is governed by the fusion rules
    \begin{eqnarray}\label{Eq:fusion}
      \Psi&\otimes& \Psi=I\nonumber\\
      \Psi&\otimes&\sigma=\sigma\nonumber\\
      \sigma&\otimes&\sigma=I\oplus\Psi,
    \end{eqnarray}
    which determine the possible resulting anyonic charges when two or more anyons in the theory are joined together. This theory is mapped to the MZM-containing system by equating the Majorana bound states with the $\sigma$ charge, the occupied state of two Majorana modes by the $\Psi$ charge, and the unoccupied state by the $\sigma$ charge. The last line of Eq.~(\ref{Eq:fusion}) therefore indicates that the combination of two Majorana modes may either contain a fermion ($\Psi$) or be empty ($I$). Importantly, any combination of anyon charges containing an odd number of $\sigma$s will always fuse to a $\sigma$, while any even combination will fuse to $I$ or $\Psi$. In the Majorana language, this means that any odd number of Majorana modes, \emph{no matter how coupled}, will always have some Majorana mode left at zero energy.

    When the experimental device shown in Fig.~\ref{Fig:Setup} is considered in these terms, it is clear that the topological case ($\epsilon_w=\eta'=0$) has a zero energy mode independent of the tuning of the dot energy $\epsilon_d$. (In this context it is useful to think of the fermionic mode on the dot being made up of two Majorana modes). When the last Majorana mode $\gamma_2$ is coupled in ($\epsilon_w,~\eta'\neq0$), the number of Majorana modes (or $\sigma$ charges) that are coupled becomes even, and an energy splitting is allowed between the states with overall even ($I$) or odd ($\Psi$) fermion parity.

    \emph{Non-zero Majorana hybridization}: If the two Majorana bound states in the nanowire have non-zero overlap, the conductance spectrum near the dot resonance becomes asymmetric in $\epsilon_d$, and the two central peaks, separated by $2\epsilon_w$ off resonance, have a crossing that approaches the dot resonance as $\epsilon_w$ increases. Such a crossing is observed in Ref.~\onlinecite{Deng16} (See Fig.~\ref{Fig:Deng}, top).

    \emph{Consequences for topological operations}: If we consider a topological operation, such as a step in a braid process \cite{Alicea11, Sau11b, Halperin11, Aasen16}, there is a constraint on the quality factor required to recover the topological result of that computation \cite{Clarke16a}. The constraint stems from the requirement that the operation must be performed diabatically with respect to the non-topological coupling $\eta'$ in order for the system to respond as if only the topologically allowed couplings are present. In terms of the quality factor introduced here, the constraint is
    \begin{equation}
      1-q\ll \frac{\hbar}{\eta\tau},
    \end{equation}
    where $\tau$ is the time taken to perform the operational step. However, in most applications there is an additional constraint that the operation be performed adiabatically with respect to the lowest excitation gap, which in this case is of the order of $\eta$ itself \cite{Nayak08,Alicea12a,Leijnse12,Beenakker2013a,Stanescu13,DasSarma15,Elliott15, Sau11b, Aasen16, Clarke16a}. Likewise, in all cases the operation must proceed quickly with respect to the hybridization $\epsilon_w$ between the two Majorana modes. This extends the requirement to
    \begin{equation}
      1-q,~\frac{\epsilon_w}{\eta}\ll \frac{\hbar}{\eta\tau}\ll 1.
    \end{equation}
    If the operation is performed too slowly compared to the quality factor and the excitation gap, or compared to the Majorana hybridization, the system will behave as if it is non-topological. If the operation is performed too quickly, diabatic errors will result \cite{Nayak08,Alicea12a,Leijnse12,Beenakker2013a,Stanescu13,DasSarma15,Elliott15,Knapp16}.
    We emphasize once more that $\epsilon_w$ and $(1-q)\eta$ are \emph{two independent energy scales} (although both are expected to be suppressed exponentially by the Majorana separation in the topological case.) It is entirely possible for a system to have a small Majorana hybridization and still behave non-topologically.

    \emph{Comparison with experimental data}:
    In Fig.~\ref{Fig:Deng}, we show data from the paper of Deng et al. measuring the conductance through a quantum dot for situations in which the Majorana hybridization is either well resolved or near zero. While we observe that a quality factor near $0.5$ seems consistent with the experimental data plotted here, these data sets were not chosen to demonstrate a high quality factor, and sample configurations with smaller splitting of the zero-bias conductance peak at resonance have been observed~\cite{Marcus15}. Note that the curvature of the outer transitions in our simple model does not match that of the experimental data, as the experimentally observed peak is bent toward the center horizon by level repulsion from the continuum states. Nevertheless, it is clearly within the realm of current experimental capabilities to perform the desired measurements, allowing direct extraction of both the Majorana hybridization and the topological quality factor from experimental data. This experiment would therefore be useful in determining whether a given system will behave topologically in more complicated experiments involving braid operations \cite{Alicea11, Sau11b, Halperin11,Clarke11b, Hyart13, Aasen16,Clarke16a} or non-local parity measurements~\cite{Plugge16, Karzig16, Vijay16}.

    \section{Acknowledgements}
    This work is supported by Microsoft, Station Q, by the Laboratory for Physical Sciences (LPS-MPO-CMTC) and by the Joint Quantum Institute (JQI-NSF-PFC). We gratefully acknowledge fruitful discussions with Jay D. Sau and Sankar Das Sarma.
\bibliography{topo-phases-1-24-17}
\end{document}